\documentclass[conference]{IEEEtran}
\IEEEoverridecommandlockouts
\usepackage{cite}
\usepackage{url}
\usepackage[utf8]{inputenc} 
\usepackage{textgreek}      

\usepackage{amsmath,amssymb,amsfonts}
\usepackage{algorithmic}
\usepackage{graphicx}
\usepackage{comment}
\usepackage{multirow} 
\usepackage{subcaption}
\usepackage{textcomp}
\usepackage{xcolor}
\def\BibTeX{{\rm B\kern-.05em{\sc i\kern-.025em b}\kern-.08em
    T\kern-.1667em\lower.7ex\hbox{E}\kern-.125emX}}
\begin{document}

\title{Optimizing Metamorphic Testing: Prioritizing Relations Through Execution Profile Dissimilarity\\
}

\author{\IEEEauthorblockN{1\textsuperscript{st} Madhusudan Srinivasan}
\IEEEauthorblockA{\textit{Computer Science Department} \\
\textit{East Carolina University}\\
Greenville, USA \\
srinivasanm23@ecu.edu}
*Corresponding author
~\\
\and
\IEEEauthorblockN{2\textsuperscript{nd} Upulee Kanewala}
\IEEEauthorblockA{\textit{Computer Science Department} \\
\textit{University of North Florida}\\
Jacksonville, USA 
}
}

\maketitle

\begin{abstract}

 An oracle is a mechanism that determines whether the output of the program for the executed test cases is correct. For machine learning programs, such an oracle is not available or is too difficult to apply. Metamorphic testing is a testing approach that uses metamorphic relations, which are necessary properties of the software under test to help verify the correctness of a program. Prioritization of metamorphic relations helps to increase fault detection effectiveness and improve metamorphic testing efficiency. However, prioritizing metamorphic relations based on faults and code coverage is often not effective for prioritizing MRs since fault-based ordering of MRs is expensive and the code coverage-based approach can provide inaccurate ordering of MRs. To this end, in this work, we propose an approach based on diversity in the execution profile of the source and follow-up test cases to prioritize metamorphic relations. We show that the proposed statement centrality-based prioritization approach increases the fault detection effectiveness by up to 31\% compared to the code coverage-based approach and reduces the time taken to detect a fault by 29\% compared to random execution of MRs. In general, our approach led to an increase in the average rate of fault detection, reduced the time taken to detect a fault, and increased fault detection effectiveness. 

\end{abstract}


\section{Introduction and Motivation}

Metamorphic testing (MT) has been applied in various domains and there is a growing need to develop techniques that allow the effective use of MT in the context of regression testing~\cite{chen2018metamorphic}. Regression testing ensures the continued functionality of software after code changes, updates, or enhancements to existing systems. As agile software development practices become increasingly prevalent, the frequency of regression testing increases with each subsequent version release~\cite{parsons2014influences}. Consequently, in such a dynamic development environment, the adoption of efficient Metamorphic Relation (MR) prioritization methods is vital. However, applying metamorphic testing in the context of regression testing faces a number of issues. First, previous research has indicated that the cost, measured in terms of time and resources, of the MT process increases proportionally to the number of MRs used~\cite{mayer2006empirical}~\cite{zhou2004metamorphic}. This is particularly pertinent because a System Under Test (SUT) can encompass MRs with multiple source and subsequent test cases. As a consequence, as the number of MRs grows, the number of test cases can grow exponentially, leading to an increase in the execution time of the MRs.

Second, applications such as machine learning and bioinformatics programs often require substantial time and resources to execute when dealing with typical data sets~\cite{fourment2008comparison}~\cite{barry2018evaluation}. Thus, prioritisation of MRs, even within a smaller MR set size, becomes an advantageous approach. Finally, research~\cite{srinivasan2018quality} has shown that individual MRs often exhibit different fault detection efficacy, and certain MRs may be geared towards identifying similar types of fault. In this context, the selection of a diverse array of MRs can significantly amplify the fault detection capabilities of MT and concurrently optimise resource utilisation during regression testing.


Prior work presented a fault and Code Coverage-based approach for prioritizing MRs in the context of regression testing~\cite{srinivasan2022metamorphic}. First, the results from applying the Code Coverage-based approach indicate that the fault detection effectiveness of the Code Coverage-based approach was equal to the random-based
approach for all the subject programs except BBMap. Also, one of the limitations of the approach is that many statements and branches in a program may not have any meaningful impact on the output. As a result, if a test covers such statements or branches, it might artificially inflate the coverage percentage without contributing to the verification of the program functionality. This reduces the effectiveness and accuracy of the prioritized MR ordering.  
Second, the Fault-based approach becomes costly when software testers have limited time to conduct metamorphic testing since the approach requires generating and executing a large number of mutants to prioritize MRs. So, we proposed a Statement Centrality-based approach to overcome these challenges.

\section{Background}
\subsection{Metamorphic Testing}
Metamorphic testing is a technique that uses metamorphic relations to detect software faults. A metamorphic relation is a mathematical or logical relationship that must hold between the inputs and outputs of a system. If the output violates this relation, it suggests a potential fault~\cite{chen2003fault}.

In metamorphic testing, a source test case (ST) generates follow-up test cases (FT) by applying a metamorphic transformation. For example, in a program that calculates the sum of a list $L = [a1, a2, ..., an]$, a source test case could be an input list $L0$ and its expected sum $sum(L0)$. A metamorphic relation for this program is that reversing $L$ to get $L'$ should not change the sum. If the sums of $L$ and $L'$ differ, it indicates a fault. This relation helps generate follow-up test cases to check the program's correctness.

\section{Proposed Approach}
\label{sec:proposed_approach}
In this approach, we propose three metrics to qualitatively analyze the dissimilarity (difference) in the execution profile of the source and follow-up test cases for an MR.
The execution profile includes information about program elements such as statements, branches, and def-use pairs that were executed by the test cases. The intuition is that the higher the dissimilarity in the execution profile between the test cases of an MR, the higher the probability of detecting failure~\cite{chen2004case}. 

Let the current version of the SUT be $v_k$. In this approach, we apply the proposed metrics in version $v_{k+1}$ and use that to prioritize the MRs for testing $v_k$ as follows: 
\begin{enumerate}
 \item Let the set of source test cases used for testing the version $v_{k}$ of the SUT  be the \emph{prioritizing source test cases} $(T_{sp})$.
    \item Let the set of follow-up test cases used for testing the version $v_{k}$ of the SUT be \emph{prioritizing follow-up test cases} $(T_{fp})$. 
    \item For each $MR_i$, log the statements (or branches) that were executed when the corresponding source and follow-up test cases from $(T_{sp})$ and $(T_{fp})$ in version $v_{k}$ of the SUT.
    \item For each $MR_i$, compute the union of statements (or branches) executed by their corresponding source and follow-up cases. Let the union of the statements be $U_{s}$.
    \item For each method in version $v_{k}$, compute the backward slice from the return statement of a method. Let the set of data and control dependent statements in backward slice be \emph{$B_{r}$}.
    \item For each statement $s \in B_{r}$, identify whether $s \in U_{s}$. If $s \notin U_{s}$, then remove s from $B_{r}$. 
   \item
       For each of the statements $s \in B_{r}$, we calculate the weight for s by applying the proposed metrics: statement affected, projected impact of the statement, and projected fault propagation. The metrics are discussed in detail in Sections \ref{label:statementaffected} to~\ref{label:faultpropogation}.
       \item Aggregate the weights of the statements to calculate a combined weight for each $MR_i$. This combined weight reflects the importance and potential impact of the MR.

\item Rank the MRs based on their combined weights. The higher the combined weight, the higher the priority of the MR.
       \item Select the top \emph{n} MRs from the prioritized ordering to execute based on the resources available for testing.
       \item Apply the prioritized MR order to test $v_{k+1}$.
       \end{enumerate}

The following subsections will explain each of the proposed metrics in detail.

\subsection{Statements Affected} \label{label:statementaffected}
\begin{figure}[h]
\centering
\includegraphics[width=0.50\textwidth]{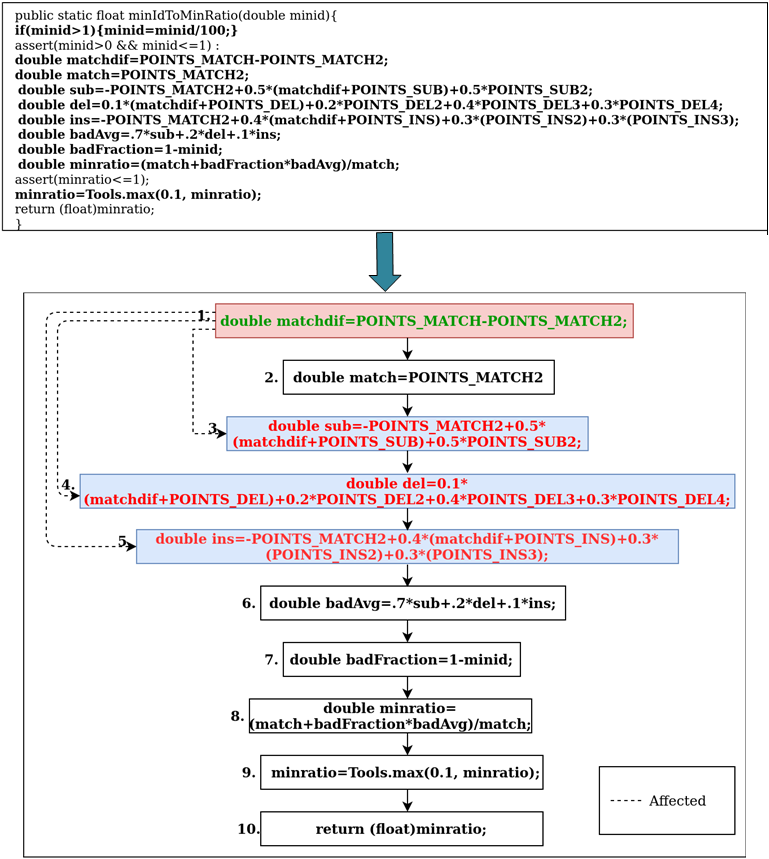}
\caption{Example to show statements affected}
\label{fig:method2_stmt_affected}
\end{figure}

The test execution can reach a faulty statement. The faulty statement causes the program to reach an incorrect program state, which is known as \emph{infection}~\cite{ammann2016introduction}. The infection in the statement propagates to all its data and control dependent statements; those statements are considered as \emph{affected statements}. In Figure~\ref{fig:method2_stmt_affected},  infection in statement 1 could affect statement 3, 4 and 5 due to a fault.  An example of statements affected is provided in Figure~\ref{fig:method2_stmt_affected}, the first node in the control flow graph (CFG) when infected due to a fault, affects nodes three, four, and five in the CFG.
To identify the affected statements for a statement s, we apply the following steps:

\begin{enumerate}
    \item For each statement $s\in B_{r}$
\begin{enumerate}
    \item Apply forward slicing to s to identify the affected statements. Let the set of statements in the forward slice of statement s be \emph{$F_{s}$}.
    \item For each statement $s \in F_{s}$, identify whether $s \in U_{s}$. If $s \notin U_{s}$, then remove s from $F_{s}$.
    \item  The total statements affected for a statement s ({$SA_{s}$}) , which is given by the formula:
\begin{equation}
\label{formula:method1_affected}
\begin{split}
\text{$SA_{s}$} = \text{$\mid F_{s} \mid$}
\end{split}
\end{equation}
where $F_{s}$ is the set of statements in forward slice.
\item Total statement affected for an MR ($TA_{s}$) is given by the formula:
\begin{equation}
\label{formula:method1_affectedMR}
\begin{split}
\text{$TA_{s}$} = \sum\limits_{s \in B_{r}} \text{$SA_{s}$}
\end{split}
\end{equation}
where $B_{r}$ is the set of statements in backward slice.
\end{enumerate}
\end{enumerate}


For example, consider the calculation of the statements affected using the example in
Figure~\ref{fig:method2_stmt_affected}. Total statements in the backward slice ($B_{r}$) is 10. Applying formula~\ref{formula:method1_affected} to statement 1, $SA_{s}$ is 5. The formula could be applied to all statements in the slice. Similarly, apply the formula~\ref{formula:method1_affectedMR} to get the total statement affected for an MR. The total statement affected for an MR is 13.
\subsection{Projected Impact of a Statement}
\label{label:nodesimpacted}
\begin{figure}[h]
\centering
\includegraphics[width=0.5\textwidth]{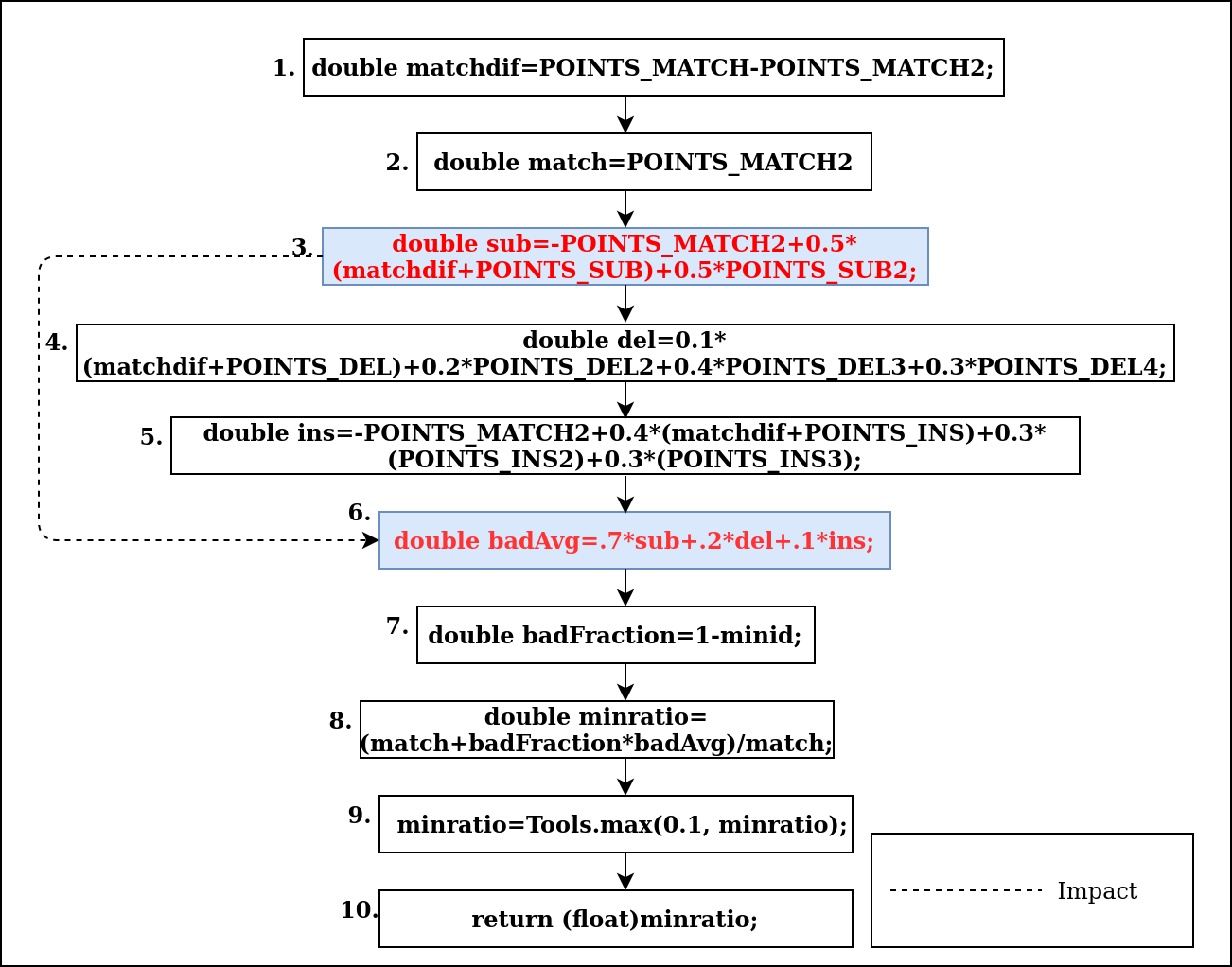}
\caption{Statement Impacted due to a affected statement}
\label{fig:statementimpact}
\end{figure}

The affected statements can propagate the infection to all its data and control dependent statements. As a result, the affected statement can have an impact or create an effect on all its dependent statements. Those statements are called \emph{impacted statement}.
For example, in Figure~\ref{fig:statementimpact}, the affected node three in the CFG impacts node six due to an infection in node 1, which is shown in the dotted line. Since node three impacts node six, the projected impact of node three is one. The projected impact of the statement is calculated using the following steps:

\begin{enumerate}
    \item For each $s\in B_{r}$
    \begin{enumerate}
    \item Compute the forward slice of s, $F_{ s}$. This provides the affected statements.
    \item For each statement s$'$ in $F_{s}$
    \begin{enumerate}
     \item Compute the forward slice $F_{ s'}$. This gives the projected impacted statement.
    \label{step1_powerimpact}
    \item For each statement $s \in F_{s'}$, identify whether $s \in U_{s}$. If $s \notin U_{s}$, then remove s from $F_{s'}$.
    \item The total projected impact of the statement ($SI_{s'}$) for a statement s$'$ is given by the formula:
    \begin{equation}
    \label{formula:impact_method2}
\begin{split}
\text{$SI_{s'}$} =\text{$\mid F_{s'} \mid$}
\end{split}
\end{equation}
 \item The total projected impact of statement ($TI_{s}$) for an MR is given by the formula:
    \begin{equation}
    \label{formula:impact_method2_MR}
\begin{split}
\text{$TI_{s}$} =\sum\limits_{s' \in F_{s} } \text{$SI_{s'}$}
\end{split}
\end{equation}
\end{enumerate}
\end{enumerate}
\end{enumerate}


For example, consider the calculation of projected impacted statements based on the example in Figure~\ref{fig:statementimpact}. The total statements in the backward slice ($B_{r}$) is 10. For statement 1 in the backward slice, the total projected statement impacted is calculated below.
\begin{enumerate}
    \item For statement 1 in the backward slice, apply the step1 to get the statement affected. The forward slice of statement 1 ($F_{s}$) is statements 3, 4, and 5.
    \item Apply step2, where we apply the forward slice ($F_{s'}$) to get the projected impact for statements 3, 4, and 5. 
    \begin{enumerate}
        \item Apply formula~\ref{formula:impact_method2}, to get the projected impact for statement 3. {$SI_{s'}$} is 1.
        \item Similarly, the projected impact for statement 4 ({$SI_{s'}$}) is 1.
        \item Similarly, the projected impact for statement 5 ({$SI_{s'}$}) is 1.
        \item The steps are carried out for all the statements in the backward slice.
        \item Apply formula~\ref{formula:impact_method2_MR}, to get the total projected impact for an MR. The total projected impact for MR ($TI_{s}$) is 27.
    \end{enumerate}
 \end{enumerate}

\subsection{Projected Fault Propagation}
\label{label:faultpropogation}
The source and follow-up test cases in an MR should execute the faulty statement and the infection caused by the fault should propagate to the output. The possibility of infection reaching the output depends on two factors: (1) the distance between the statement in the program to the output and (2) the total number of operations in the statement. In our work, the return statement is considered the output statement since the result of a specific task performed by a method is passed to the return statement.

In this metric, we use the two factors mentioned earlier to calculate the fault propagation ability of statements in the backward slice. The distance is greater between the statement and output; then, it is less likely that the infection would reach the output statement. The number of operations in each statement in the backward slice is taken into consideration, since more arithmetic and conditional operations in a statement increase the likelihood of error in the program. The steps to calculate the propagation of the fault from an impacted statement are provided below.


\begin{enumerate}
\item For each statement $s \in B_{r}$
        \begin{enumerate}
        \item \emph{Generate abstract syntax tree} (AST) for s. The AST is used to count the number of arithmetic or conditional operators in s.  \label{step2_faultpropogation}
   \label{step3_faultpropogation}
    \item Traverse from s to the return statement in the control flow graph (CFG) to get the distance between s to output. \label{step6_faultpropogation}
    \item 
    We calculate the projected fault propagation from statement $s$ ($PF_{s}$) to the output using the formula.
\begin{equation}
\label{formula:faultprop_method2}
PF_{s} = \frac{1}{\text{Total operations in } s + \text{Distance between } s \text{ and output}}
\end{equation}

Where,
Total Operations in s= Total Operators in AST for s, 

Distance between s and output= Number of hops from s to the output node (return statement) in CFG.
\item
We calculate the total fault propagation ($TFP$) for an MR using the formula.
\begin{equation}
\label{formual:faultprop_method2_MR}
\begin{split}
\text{$TFP$} = \sum\limits_{s \in B_{r} } \text{$PF_{s}$}
\end{split}
\end{equation}
\end{enumerate}
\end{enumerate}


For example, consider the calculation of projected fault propagation based on the example in Figure~\ref{fig:statementimpact}. Statement 3 impacts statement 6. The number of operations in statement 6 is 5. The distance between statement 6 and output statement 10 is 4. We apply formula~\ref{formula:faultprop_method2} to get the projected fault propagation for statement 6. The projected fault propagation ($PF_{s}$) for statement 6 is 0.1. Total fault propagation (TFP) for an MR using formula~\ref{formual:faultprop_method2_MR} is 10.6.

\subsection{MR Quality Score}
 The proposed metrics indicate the degree of dissimilarity between the execution profile of the test cases and help to determine the MR quality score. The MRs are prioritized using the MR quality score. The score is based on the total statements affected, the total projected impact of the statements, and the projected fault propagation.
The quality score for an MR is calculated using the formula:
  \begin{equation}
\begin{split}
\text{MR Quality Score} = \text{Total Statements affected  ($TA_{s}$)} \\+  \text{Total Projected Impact of Statement ($TI_{s}$)} \\+ \text{Total Projected Fault Propagation (TFP)}.
\end{split}
\end{equation}

\section{Evaluation}
In this work, we plan to evaluate the utility of the developed prioritization approaches similar to  work~\cite{srinivasan2022metamorphic} and on the following aspects: (1) fault detection effectiveness of MR sets, (2) effective number of MRs required for testing, (3) time taken to detect a fault and (4) average percentage of faults detected. We evaluate the effectiveness of the Statement Centrality-based prioritizing approaches with (1) a \emph{random baseline}: this represents the current practice of executing source and follow-up test cases of MRs in random order. (2) \emph{Fault-based ordering}: this represents the MR order based on the fault detection capability of the MRs. The approach selects an MR that has killed the highest number of mutants and places it in the prioritized MR ordering. The prioritization process continues until all MRs are prioritized and faults have been exposed~\cite{srinivasan2022metamorphic}. (3) \emph{Code Coverage-based ordering}: this represents the MR order based on the statement and branch-coverage information of the MRs. The approach selects the MR which has covered the highest statements or branches in the SUT and places it in the prioritized MR ordering. The prioritization continues until all statements and branches are covered~\cite{srinivasan2022metamorphic}. 

We conducted experiments to find answers to the following research questions:
\begin{enumerate}
    \item Research Question 1 (RQ1): \emph{How do the proposed MR prioritization approach perform against the random baseline?}
    \item Research Question 2 (RQ2): \emph{How do the Statement Centrality-based approaches perform against statement and branch coverage-based prioritization approaches?}
      \item Research Question 3 (RQ3): \emph{How do the Statement Centrality-based approach perform against the Fault-based approach?}
    \end{enumerate}

\subsection{Evaluation Procedure}
\label{sec:evalproc_method2}
In order to answer the above research questions, we carry out the following validation procedure similar to the work~\cite{srinivasan2022metamorphic}:

\begin{enumerate}
\item We create a set of source test cases for the SUT independent of the $T_{sp}$ for validation. These source test cases will be called \emph{validation source test cases} $(T_{sv})$. Then we create the follow-up test cases according to the MRs used for conducting MT on a given SUT. These will be called \emph{validation follow-up test cases} ($T_{fv}$). 

\item We generate a set of mutants for the version $v_k$ of the SUT. These mutants will be referred to as the \emph{validation set of faults} $(F_v)$. Then from the generated mutants, we remove mutants that give exceptions or mutants that do not finish execution from further consideration.

\item We use the method described in Section~\ref{sec:proposed_approach} to obtain Statement Centrality-based MR ordering. Then we applied the obtained MR ordering to $F_v$ and recorded the mutant killing information.

\item \textit{Creating the random baseline}: We generate 100 random MR orderings and apply each of those orderings to $F_v$. The mutant killing information for each of those random orders is recorded. The average mutant killing rates of these 100 random orderings were computed to obtain the fault detection effectiveness of the random baseline.
\item \textit{Creating the Fault-based ordering}: 
\begin{enumerate}

\item Let the set of faults detected in version $v_{k-1}$ be \emph{prioritizing set of faults} $F_p$. For each fault $f \in F_p$, log whether each $MR_i$ revealed $f$ when executed with $(T_{sp})$ and $(T_{fp})$. Here, $MR_i$ refers to metamorphic relations ($MR_1$, $MR_2$....$MR_n$) used for testing with i=1 to n.
\item Apply the greedy approach to create the prioritized MR ordering as follows~\cite{srinivasan2022metamorphic}:
\begin{enumerate}
\item Select the MR that has revealed highest number of faults and place it in the prioritized MR ordering.
\item Remove the faults revealed by the MR.
\item Repeat the previous two steps until all faults are covered.    
\end{enumerate}
\end{enumerate}
\item \textit{Creating the Code Coverage based ordering}: 
\begin{enumerate}
\item For each $MR_i$, log the statements (or branches) that were executed when running the corresponding source and follow-up test cases from $(T_{sp})$ and $(T_{fp})$ on version $v_{k-1}$ of the SUT. Here, $MR_i$ refers to metamorphic relations ($MR_1$, $MR_2$....$MR_n$).
\item For each $MR_i$, compute the union of statements (or branches) executed by their corresponding source and follow-up cases.
\item Use the following greedy approach based on the statement (or branch) coverage to create the prioritized ordering of the MRs for testing version $v_k$.
    \begin{enumerate}
\item Select the MR that covers the highest number of statements (or branches)
and place it in the prioritized MR ordering. If there are multiple MRs with
the same highest coverage, select one MR from them randomly.
\item Remove the statements (or branches) covered by that MR from further consideration.
\item Repeat the previous two steps until all the all the possible statements (or branches) are covered.
\end{enumerate}
\end{enumerate}

\end{enumerate}

\subsection{Evaluation Measures}
\label{evalMeasures}
We used the following measures to evaluate the effectiveness of the MR orderings generated by our MR prioritization approaches:
\begin{enumerate}
    \item To measure the \emph{fault detection effectiveness} of a given set of MRs, we use the percentage of mutants killed by those MRs. 
    \item To calculate the \emph{effective MR set size}, we used the following approach: The fault detection effectiveness of MT typically increases with the number of MRs used for testing. However, beyond a certain point, this increase slows due to MR redundancy. When the increase in fault detection between MR sets of sizes $m$ and $m+1$ (where $m+1$ is created by adding an MR to $m$) is insignificant, the effective MR set size can be determined. Specifically, if the difference in fault detection effectiveness between MR sets of sizes $m$ and $m+1$ is below a threshold, $m$ is considered the effective MR set size. The threshold value should consider the critical nature of the SUT. In this work, we used thresholds of 5\% and 2.5\%, as used in previous related work~\cite{gay2015automated}.
    \item We used the following approach to find the \emph{average time taken to detect a fault}: For each killable mutant $m$ in $F_v$, we calculated the time taken to kill the mutant ($t_m$) by computing the time to execute the source and follow-up test cases of the MRs in a given prioritized order until $m$ is killed. The average time to detect a fault is then computed using the following formula:

    \begin{equation}
\begin{split}
 \frac{\sum t_m}{\text{\# killable mutants in $F_v$}}
\end{split}
\end{equation}

     \item We use a metric called \emph{average percentage of fault detected} (APFD) developed by Elbaum et al.~\cite{elbaum2002test}~\cite{malishevsky2006cost}~\cite{elbaum2004selecting} that measures the average rate of fault detection per percentage of test suite execution. The APFD is calculated by taking the weighted average percentage of faults detected by the MRs. APFD can be calculated using the formula below: 
    \begin{equation}
\begin{split}
\text{APFD} =
 \frac{1-MRF_1+MRF_2+...+MRF_m}{nm}+\frac{1}{2n}
\end{split}
\label{label:apfd_formula}
\end{equation}
Where MR represents the metamorphic relations under evaluation, m represents the number of faults present in the SUT and n represents the total number of MRs used.

\end{enumerate}

\subsection{Subject Programs and MRs}
    
   We applied the proposed approach to three applications from different domains to evaluate our MR prioritization methods. These projects include BBMap (genetics), LingPipe (computational linguistics and bioinformatics), and machine learning models (linear regression and naive Bayes). This diversity ensures that our MR prioritization methods are generalizable across various fields. Additionally, these projects vary in complexity and handle large datasets, making them ideal for evaluating our approach.
 \begin{itemize}
\item \textbf{BBMap}\footnote{\url{https://jgi.doe.gov/data-and-tools/bbtools/bb-tools-user-guide/BBMap-guide/}}: Aligns RNA and DNA sequence reads to a reference genome. Inputs are a set of reads and a reference genome, and the output is the mapping of reads to the genome. BBMap consists of 100,000 LOC.
\item \textbf{LingPipe}\footnote{\url{http://alias-i.com/}}: Processes text using computational linguistics, often for bio-entity recognition in bioinformatics. Inputs are bio-medical articles, and the output is the extracted bio-entities. LingPipe has 50,000 LOC.
\item \textbf{Linear Regression}\footnote{\url{https://weka.sourceforge.io/doc.dev/weka/classifiers/functions/LinearRegression}}: Models the relationship between input variables and a single output variable. Implemented in Weka as the \textit{LinearRegression()} class with 500 LOC. Inputs are training and test datasets in .arff format, and the output is the prediction on test data.
\item \textbf{Naive Bayes}\footnote{\url{https://weka.sourceforge.io/doc.dev/weka/classifiers/bayes/NaiveBayes.html}}: A probabilistic classifier used for classification tasks, implemented in Weka as the \textit{NaiveBayes()} class with 500 LOC. Inputs are training and test datasets in .arff format, and the output is the classification on test data.
\end{itemize}

\subsection{Metamorphic Relations}
To conduct MT on the above-mentioned subject programs, we used a set of MRs developed in previous studies. For testing BBMap, we used eight MRs developed by Giannoulatou et al. based on the expected behavior of short-read alignment software~\cite{giannoulatou2014verification}.  All of these MRs specify modifications to the reads supplied as input, such as randomly shuffling the reads, duplicating reads, and removing reads. 

For conducting MT on IBk, we used 11 MRs developed by Xie et al.~\cite{xie2011testing}. These MRs are developed on the basis of the user expectations of supervised classifiers. These MRs modify the training and testing data so that the predictions do not change between the source and follow-up test cases.   

To test the bio-entity recognition task in LingPipe, we used ten MRs developed in work~\cite{srinivasan2018quality}. We describe three categories of MRs for testing bio-entity recognition software. These MR categories are addition, deletion, and shuffling. In addition to relations, we extend the text by adding new text such that the sentence/paragraph boundaries are preserved. In the deletion relation, the text is truncated by removing part of it so that the sentence boundaries are preserved. In the shuffling relations, we shuffle portions of the text. Similarly, to test the linear regression system, we used 11 MRs developed by Luu et al.~\cite{luu2021testing}. The properties of the linear regression are related to the addition of data points, the rescaling of inputs, the shifting of variables, the reordering of data, and the rotation of independent variables. These MRs were grouped into 6 categories and derived from the properties of the targeted algorithm.

\subsection{Source and Follow-up Test Cases}
\label{testcases}

As we described before, MT involves the process of generating source and follow-up test cases based on the MRs used for testing. BBMap requires a set of reads and a reference genome as inputs. We used the E.coli reference genome and a set of its reads and Yeast reference genome and a set of its reads~\footnote{http://genome.ucsc.edu/} as source test cases. 
E.coli and Yeast are considered model genomes and are widely used in the bioinformatics domain to conduct tests. To generate the follow-up test cases using these source test cases, we applied the input transformations described in the MRs. 

IBk uses a training data set to train the k-nearest neighbor classifier, and a test data set is used to evaluate the performance of the trained classifier. 
After executing this source test case, the output will be the class labels predicted for the test data set, which is a value from the set \{0,1,2,3,4\} in this example. We generated ten similar datasets randomly, where each training and test data set contained four numerical attributes and a class label. The attribute values are randomly selected within the range [0, 100]. The values of the class label are randomly selected from the set \{0,1,2,3,4,5\}. The size of the training testing data sets ranges within [0, 200]. To generate the follow-up test cases using these source test cases, we applied the input transformations described in the MRs. 

For creating source test cases for LingPipe, two biomedical articles with the PMCIDs 100320 and 100325 were obtained through PubMed~\footnote{https://www.ncbi.nlm.nih.gov/pubmed}. 
The sentences and paragraphs used as source test cases for the MRs were randomly selected from these two articles. However, for specific MRs related to removing some words from a list of random words (MR8) and shuffling a list of random words (MR10), 15 random articles were selected and two sets of 500 random words were chosen from the selected articles to generate the source test case. To generate the follow-up test cases using these source test cases, we applied the input transformations described in the MRs. 

Linear Regression uses a training data set to train the model, and a test data set is used to evaluate the performance of the trained model. We obtained training data from machine learning repositories \footnote{https://archive.ics.uci.edu/ml/datasets/YearPredictionMSD},\footnote{https://archive.ics.uci.edu/ml/datasets/Superconductivty+Data} as the source data sets. The first data set contains 515345 instances and 90 attributes. The second data set contains 21263 instances and 81 attributes. To generate the follow-up data sets using these source data sets, we applied the input transformations described in the MRs. Similarly, for Naive Bayes, we obtained training data from machine learning repositories \footnote{https://archive.ics.uci.edu/ml/datasets/Adult},\footnote{https://www.kaggle.com/datasets/rakeshrau/social-network-ads} as source data sets. The first data set contains 48842 instances and 14 attributes. The second data set contains 400 instances and 5 attributes. To generate the follow-up data set using this source data set, we applied the input transformations described in the MRs. 

\subsection{Mutant Generation}
\label{sec:mutants}
For each version of the subject program, we aimed at developing generated mutant set to be used as $F_v$. For this, we used three automated mutation tools: $\mu$Java\footnote{https://github.com/jeffoffutt/muJava}, PIT\footnote{http://pitest.org/} and Major\footnote{http://mutation-testing.org/}. For generating mutants using $\mu$Java, we used all the method level mutation operators~\cite{ma2006mujava}.  With the PIT mutation tool~\cite{coles2016pit} and the Major mutation tool~\cite{just2014major} we used all the available mutation operators provided by the tools. All mutants generated using these tools were \emph{first order mutants}, where a mutation operator is used to make a single modification to the source code to generate the mutant.  Table~\ref{Tab:mutants_method2} shows the number of mutants used for evaluation. We used two mutation testing tools for each SUT, while tools that were not used are represented as N/A.
 

\begin{table}[h]
\centering
\caption{Mutants generated for version $v_{k}$ of the SUTs}
\label{Tab:mutants_method2}
\begin{tabular}{|c|c|c|c|c|}
\hline
\textbf{Version} & \textbf{Subject} & \textbf{\#Major} & \textbf{\#PIT} & \textbf{\# $\mathbf{\mu}$Java} \\ \hline
38.71            & BBMap            & 102              & 100            & N/A             \\ \hline
3.8.6            & Linear Regression         & 377              &   73         & N/A             \\ \hline
3.8.6            & Naive Bayes         & 452              & 452            & N/A             \\ \hline

4.1.0            & LingPipe         & 100              & N/A            & 100             \\ \hline
\end{tabular}
\end{table}



\begin{table*}[h]
    \centering
    \caption{Validation Setup}
    \small\label{table:method2_evalsetup_tests}
    \begin{tabular}{|c|c|c|c|c|}
        \hline
        \textbf{Subject} & \textit{$\mathbf{T_{sp}}$} & \textit{\textbf{SUT Version}} & \textit{$\mathbf{T_{sv}}$} & \textit{$\mathbf{F_v}$} \\ \hline
        \multirow{2}{*}{BBMap} & Ecoli & 38.70 & Yeast & PIT (38.71) \\ \cline{2-5} 
         & Yeast & 38.70 & Ecoli & Major (38.71) \\ \hline \hline
        \multirow{2}{*}{LingPipe} & ID:100325 & 3.9.3 & ID:100320 & PIT (4.1.0) \\ \cline{2-5} 
         & ID:100320 & 3.9.3 & ID:100325 & PIT (4.1.0) \\ \hline \hline
        \multirow{2}{*}{Linear Regression} & YearPrediction & 3.8.5 & SuperConduct & Major (38.71) \\ \cline{2-5} 
         & SuperConduct & 3.8.6 & YearPrediction & PIT (4.1.0) \\ \hline \hline
        \multirow{2}{*}{Naive Bayes} & Adult & 3.8.5 & SocialNetwork & PIT (4.1.0) \\ \cline{2-5} 
         & SocialNetwork & 3.8.6 & Adult & Major (38.71) \\
        \hline
    \end{tabular}
\end{table*}







\section{Results and Analysis}

\label{section:Results&Discussion_method2}
\begin{figure*}[h]
\centering
  \begin{subfigure}[b]{0.49\textwidth}
    \includegraphics[width=\textwidth,height=5.5cm]{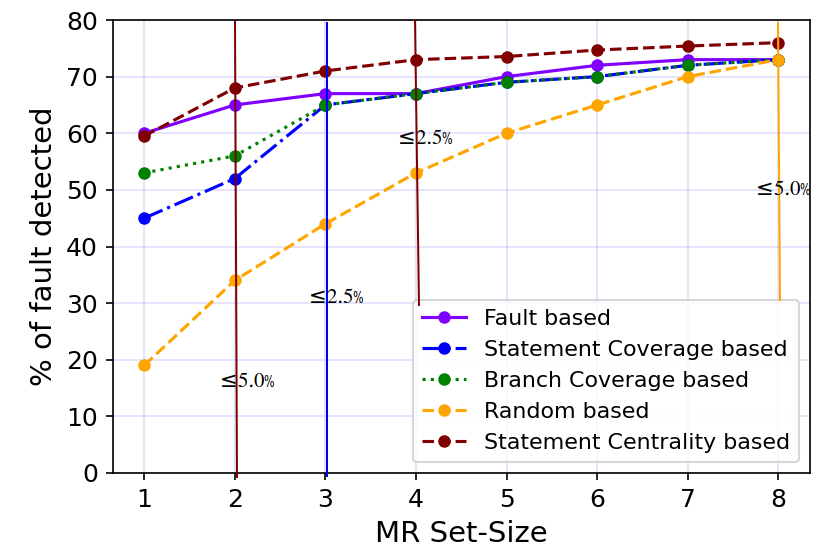}
    \caption{BBMap}
    \label{fig:BBMap_method2_killrateMRs}
  \end{subfigure}
  \hfill
\begin{subfigure}[b]{0.49\textwidth}
      \includegraphics[width=\textwidth,height=5.5cm]{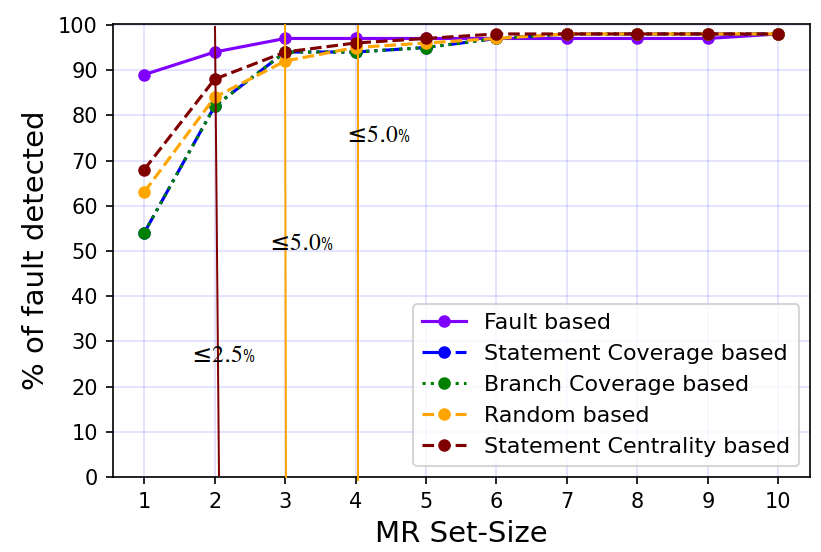}
    \caption{LingPipe}
    \label{fig:LingPipe_method2_killrateMRs}
  \end{subfigure}
    \hfill
   \begin{subfigure}[b]{0.49\textwidth}
      \includegraphics[width=\textwidth,height=5.5cm]{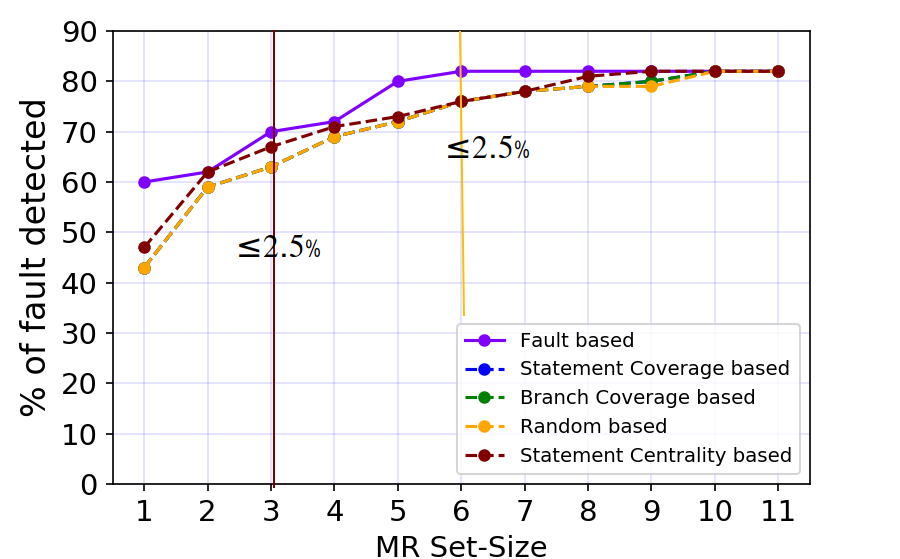}
    \caption{Linear Regression}
    \label{fig:LR_method2_killrateMRs}
  \end{subfigure}
    \hfill
     \begin{subfigure}[b]{0.49\textwidth}
      \includegraphics[width=\textwidth,height=5.5cm]{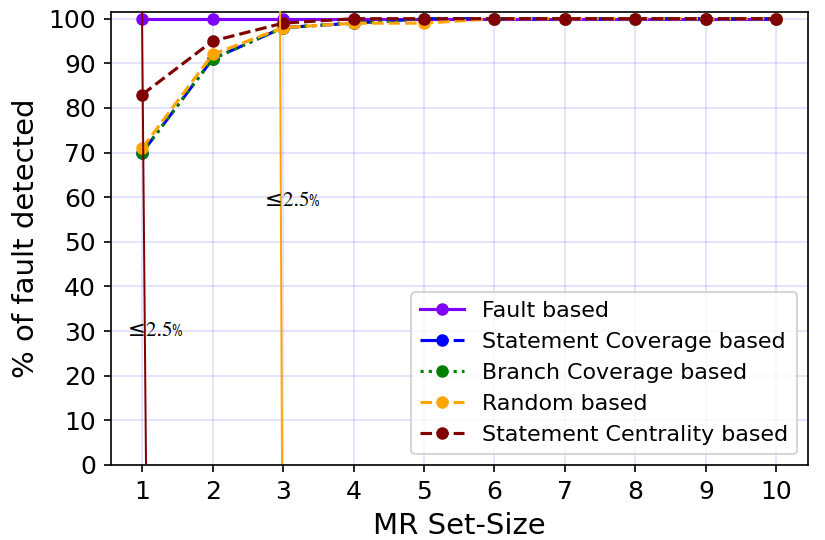}
    \caption{Naive Bayes}
    \label{fig:NB_method2_killrateMRs}
  \end{subfigure}
        \caption{Fault detection effectiveness for BBMap, LingPipe, Linear Regression and Naive Bayes}
    \label{Fig:subject_method2_killrateMRs}
\end{figure*}

In this Section, we discuss our experimental results and provide answers to the three research questions that we listed in Section~\ref{sec:evalproc_method2}. For each subject program, we performed the validation procedure described in Section~\ref{sec:evalproc_method2} using the setup described in Table~\ref{table:method2_evalsetup_tests}. In this setup, we used the generated mutant sets and the source test cases as $T_{sp}, F_v,$ and ${T_{sv}}$ in two different configurations. For example, the first evaluation run for BBMap was conducted by executing Ecoli as $T_{sp}$ on version 38.70 ($v_{k-1}$), Yeast as $T_{sv}$, and PIT mutants generated on version 38.71 ($v_k$) as $F_{v}$ (refers to row 1 of Table~\ref{table:method2_evalsetup_tests}). The second evaluation run for BBMap was conducted by executing Yeast as $T_{sp}$ on version 38.70 ($v_{k-1}$), Ecoli as $T_{sv}$, and Major mutants generated on version 38.71 ($v_k$) as $F_{v}$. In Table~\ref{table:method2_evalsetup_tests} for IBk, Test1 refers to 5 of the randomly generated datasets and Test2 refers to the other five datasets. Figure~\ref{Fig:subject_method2_killrateMRs} shows the average fault detection effectiveness for evaluation runs described above vs. the MR set size used for testing, for each subject program. We also plot the percentage of faults detected by the random baseline, Fault-based ordering and Code Coverage-based MR ordering for comparison. Table~\ref{tab:stmt_random}, Table~\ref{tab:stmtcentrality_smtcov_subjects} and Table~\ref{tab:stmtcentral_faultbased_subjects} provides the relative fault detection effectiveness for SUT. We apply the paired permutation test to each MR set size for each of the subject programs with α= 0.05, and the relative improvements that are significant are marked with a. We presented the hypothesis for each research question a) $H_{01}$: statement centrality approach (SC) is greater than the random approach in fault detection for MR sets of size m. b) $H_{02}$: SC is greater than statement and branch coverage for MR sets of size m. c) $H_{03}$: SC is greater than fault based ordering in fault detection for MR sets of size m. The null hypothesis $H_{0x}$ for each of the above defined
hypothesis $H_{X}$ is that the Statement Centrality-based approach
performs equal to or worse than the random baseline, code coverage or the fault based ordering.

\subsection{RQ1: Comparison of MR Prioritization Approaches versus Random Baseline}

\subsubsection{Fault detection effectiveness}


In Table~\ref{tab:stmt_random}, we list the relative improvement in average fault detection effectiveness for the two MR prioritization methods over the currently used random approach for BBMap, LingPipe, Linear Regression, and Naive Bayes respectively. As shown in Table~\ref{tab:stmt_random} for BBMap, the Statement Centrality-based approach showed significant improvements in the average fault detection effectiveness over the random approach for all MR set sizes except for the last set size. Particularly, among the different MR set sizes, the increase in fault detection percentage varies from 4.10\% to 215.78\% between the two methods. Similarly for LingPipe, our proposed approaches show improvements for MR set sizes $m=1$ to $m=6$. 

Furthermore, overall, the relative improvement across the MR prioritization methods varies between 0\% - 7\%. For Linear Regression our proposed approaches show improvement over the random approach for all the MR set sizes. Similarly, for Naive Bayes, our proposed approaches show improvement in fault detection effectiveness for MR set size $m=1$ to $m=5$. So, we reject the null hypothesis in general. \textbf{Therefore, our proposed approach shows improvement in fault detection effectiveness over random based prioritization for all our subject programs.}

\begin{table}[h]
\centering
 \caption{Relative improvement in fault detection effectiveness of Statement Centrality-based approach compared to random baseline for subject programs}
 \label{tab:stmt_random}
\begin{tabular}{|p{1.2cm}|p{1.2cm} p{1.2cm}|p{1.2cm}|p{1.2cm}}
\hline
\textbf{MR Set Size} & \multicolumn{1}{c|}{\textbf{BBMap}} & \textbf{LingPipe} & \textbf{Linear Regression} & \multicolumn{1}{c|}{\textbf{Naive Bayes}} \\ \hline
1                    & \multicolumn{1}{c|}{215.78\%*}         & 7.93\%*              & 2.17\%*                       & \multicolumn{1}{c|}{16.40\%*}                \\ \hline
2                    & \multicolumn{1}{c|}{100\%*}            & 4.76\%*              & 5.08\%*                       & \multicolumn{1}{c|}{2.70\%*}                 \\ \hline
3                    & \multicolumn{1}{c|}{59.09\%*}          & 2.17\%*              & 3.07\%*                       & \multicolumn{1}{c|}{0.81\%}                 \\ \hline
4                    & \multicolumn{1}{c|}{35.84\%*}          & 1.05\%*              & 2.89\%*                       & \multicolumn{1}{c|}{1.01\%*}                 \\ \hline
5                    & \multicolumn{1}{c|}{17.74\%*}          & 1.04\%             & 1.38\%                       & \multicolumn{1}{c|}{0.41\%}                 \\ \hline
6                    & \multicolumn{1}{c|}{16.92\%*}          & 1.03\%*              & 2.70\%*                       & \multicolumn{1}{c|}{0\%}                    \\ \hline
7                    & \multicolumn{1}{c|}{4.10\%*}           & 0\%                 & 2.63\%*                       & \multicolumn{1}{c|}{0\%}                    \\ \hline
8                    & \multicolumn{1}{c|}{0\%}              & 0\%                 & 5.19\%*                       & \multicolumn{1}{c|}{0\%}                    \\ \hline
9                    & \multicolumn{1}{c|}{}               & 0\%                 & 5.12\%*                       & \multicolumn{1}{c|}{0\%}                    \\ \cline{1-1} \cline{3-5} 
10                   & \multicolumn{1}{c|}{}               & 0\%                 & 2.5\%*                        & \multicolumn{1}{c|}{0\%}                    \\ \cline{1-1} \cline{3-5} 
11                   &                                     &                   & 0\%                          &                                           \\ \cline{1-1} \cline{4-4}
\end{tabular}
\end{table}



\subsubsection{Effective number of MRs used for Testing}
In Figures~\ref{fig:BBMap_method2_killrateMRs},~\ref{fig:LingPipe_method2_killrateMRs},~\ref{fig:LR_method2_killrateMRs} and~\ref{fig:NB_method2_killrateMRs}, we show the effective number of MRs using the fault detection thresholds for BBMap, LingPipe, Linear Regression and Naive Bayes respectively. The brown vertical lines represent the effective size of the MR set when using the Statement Centrality-based prioritization and the orange vertical lines represent the same for the random baseline. The respective thresholds are shown near each vertical line.

As shown in Figure~\ref{fig:BBMap_method2_killrateMRs}, for BBMap, the effective MR set size is 2 for the 5\% threshold when statement-centrality based prioritization is used. With the random baseline, the effective MR set size is 8. For LingPipe, the effective MR set size is two for the 2. 5\% threshold when using a statement centrality-based approach. With the random baseline, the 5.0\% thresholds are only met with MR set size 3 and the 2.5\% threshold is met with MR set size 6. Similarly, for linear regression, the MR set size is 4 for the 2.5\% threshold when the Statement Centrality-based approach is used. Compared to the random baseline and code coverage, the size of the MR set is 6 for the 2.5\% threshold. For Naive Bayes, the MR set size is 3 is met at 2. 5\% when Statement Centrality-based is used. With a random baseline, MR set size is 3 for a threshold of 2. 5\%.  \textbf{Therefore for all the subject programs, using Statement Centrality-based prioritization resulted in a reduction in the effective MR set size compared to using a random and Code Coverage-based ordering of MRs.}

\subsubsection{The average time taken to detect a fault } Tables ~\ref{time_method_SUT} show the average time taken to detect a fault for the subject programs. As shown in these results, the use of \textbf{ Statement Centrality-based prioritization resulted in reductions in the average time taken to detect a fault ranging from 39\%-228\% relative to the random baseline for the SUT}. 
\subsubsection{Average percentage of fault detected} Table~\ref{tab:apfd_subjects} shows the APFD for the SUT. We can observe that for all our subject programs, our proposed approach provided a higher APFD value between 0.87-0.97 and shows an average 15\% increase in fault detection rate compared to random-based prioritization. \textbf{As a result, our approach provides a higher fault detection rate compared to the random approach.} \textbf{ Answer RQ1: Our proposed approach outperformed the random-based approach for all our subject programs.}

\begin{table}[h]
\centering
\caption{Comparison of average time taken to detect a fault using Statement Centrality-based, fault based, random and code coverage based approaches}
 \label{time_method_SUT}
\begin{tabular}{|p{1.3cm}|p{1.3cm}|p{1.3cm}|p{1.3cm}|p{1.3cm}|}

\hline
\textbf{Subject}  & \textbf{Statement Centrality-based} & \textbf{Fault based} & \textbf{Random based} & \textbf{Code Coverage based} \\ \hline
BBMap             & 75s                                 & 108s                 & 230s                  & 145s                         \\ \hline
LingPipe          & 108s                                & 100s                 & 180s                  & 120s                         \\ \hline
Linear Regression & 127s                                & 84s                  & 153s                  & 153s                         \\ \hline
Naive Bayes       & 110s                                & 97s                  & 153s                  & 150s                         \\ \hline
\end{tabular}
\end{table}

\begin{table}[h]
\centering
\caption{Average percentage of fault detected for subject programs}
\label{tab:apfd_subjects}
\begin{tabular}{|p{1.3cm}|p{1.3cm}|p{1.3cm}|p{1.3cm}|p{1.3cm}|}
\hline
\textbf{Subject}  & \textbf{Statement Centrality-based} & \textbf{Fault based} & \textbf{Random based} & \textbf{Code Coverage based} \\ \hline
BBMap             & 0.91                                & 0.98                 & 0.65                  & 0.75                         \\ \hline
LingPipe          & 0.87                                & 0.90                 & 0.76                  & 0.73                         \\ \hline
Linear Regression & 0.92                                & 0.97                 & 0.83                  & 0.83                         \\ \hline
Naive Bayes       & 0.90                                & 0.95                 & 0.88                  & 0.87                         \\ \hline
\end{tabular}
\end{table}

\subsection{RQ2: Comparison of Statement Centrality-based prioritization approach against the coverage-based prioritization approaches}

\begin{table}[h!]
\centering
\caption{Average relative improvement in fault detection effectiveness using statement-centrality based approach over Statement coverage-based prioritization approaches}
\label{tab:stmtcentrality_smtcov_subjects}

\begin{tabular}{|p{1.2cm}|p{1.2cm} p{1.2cm}|p{1.2cm}|p{1.2cm}}
\hline
\textbf{MR Set-Size} & \multicolumn{1}{c|}{\textbf{BBMap}} & \textbf{LingPipe} & \textbf{Linear Regression} & \multicolumn{1}{c|}{\textbf{Naive Bayes}} \\ \hline
1                    & \multicolumn{1}{c|}{31.33\%*}          & 26.18\%*           & 4.44\%*                       & \multicolumn{1}{c|}{19.08\%*}                \\ \hline
2                    & \multicolumn{1}{c|}{30.76\%*}          & 14.67\%*          & 5.08\%*                       & \multicolumn{1}{c|}{3.93\%*}                 \\ \hline
3                    & \multicolumn{1}{c|}{7.69\%*}           & 6.27\%*           & 3.07\%*                       & \multicolumn{1}{c|}{1.02\%*}                 \\ \hline
4                    & \multicolumn{1}{c|}{7.46\%*}           & 6.08\%*           & 2.89\%*                       & \multicolumn{1}{c|}{0.90\%}                 \\ \hline
5                    & \multicolumn{1}{c|}{5.79\%*}           & 4.61\%*           & 1.38\%                       & \multicolumn{1}{c|}{0.30\%}                 \\ \hline
6                    & \multicolumn{1}{c|}{8.57\%*}           & 2.92\%*           & 2.70\%*                       & \multicolumn{1}{c|}{0\%}                    \\ \hline
7                    & \multicolumn{1}{c|}{5.55\%*}           & 2.67\%*           & 2.63\%*                       & \multicolumn{1}{c|}{0\%}                    \\ \hline
8                    & \multicolumn{1}{c|}{2.74\%*}           & 2.74\%            & 5.19\%*                       & \multicolumn{1}{c|}{0\%}                    \\ \hline
9                    & \multicolumn{1}{c|}{}               & 0                 & 3.79\%*                       & \multicolumn{1}{c|}{0\%}                    \\ \cline{1-1} \cline{3-5} 
10                   & \multicolumn{1}{c|}{}               & 0                 & 2.5\%*                        & \multicolumn{1}{c|}{0\%}                    \\ \cline{1-1} \cline{3-5} 
11                   &                                     &                   & 0                          &                                           \\ \cline{1-1} \cline{4-4}
\end{tabular}
\end{table}

\begin{table}[h!]
\centering
\caption{Average relative improvement in fault detection effectiveness using statement-centrality based approach over branch coverage-based prioritization approaches}
\label{tab:stmtcentrality_branchcov_subjects}
\begin{tabular}{|p{1.2cm}|p{1.2cm} p{1.2cm}|p{1.2cm}|p{1.2cm}}
\hline
\textbf{MR Set-Size} & \multicolumn{1}{c|}{\textbf{BBMap}} & \textbf{LingPipe} & \textbf{Linear Regression} & \multicolumn{1}{c|}{\textbf{Naive Bayes}} \\ \hline
1                    & \multicolumn{1}{c|}{13.33\%*}        & 26.18\%*           & 4.44\%*                      & \multicolumn{1}{c|}{19.08\%*}                \\ \hline
2                    & \multicolumn{1}{c|}{21.42\%*}        & 14.67\%*          & 5.08\%*                       & \multicolumn{1}{c|}{3.93\%*}                 \\ \hline
3                    & \multicolumn{1}{c|}{7.69\%*}           & 6.27\%*           & 3.07\%*                       & \multicolumn{1}{c|}{1.02\%*}                 \\ \hline
4                    & \multicolumn{1}{c|}{7.46\%*}           & 6.08\%*           & 2.89\%*                       & \multicolumn{1}{c|}{0.90\%}                 \\ \hline
5                    & \multicolumn{1}{c|}{5.79\%*}           & 4.61\%*           & 1.38\%                       & \multicolumn{1}{c|}{0.30\%}                 \\ \hline
6                    & \multicolumn{1}{c|}{8.57\%*}           & 2.92\%*           & 2.70\%*                      & \multicolumn{1}{c|}{0\%}                    \\ \hline
7                    & \multicolumn{1}{c|}{5.55\%*}           & 2.67\%*           & 2.63\%*                       & \multicolumn{1}{c|}{0\%}                    \\ \hline
8                    & \multicolumn{1}{c|}{0\%}            & 2.74\%*            & 5.19\%*                       & \multicolumn{1}{c|}{0\%}                    \\ \hline
9                    & \multicolumn{1}{c|}{}               & 0\%                 & 3.79\%*                       & \multicolumn{1}{c|}{0\%}                    \\ \cline{1-1} \cline{3-5} 
10                   & \multicolumn{1}{c|}{}               & 0                 & 2.5\%*                        & \multicolumn{1}{c|}{0\%}                    \\ \cline{1-1} \cline{3-5} 
11                   &                                     &                   & 0\%                          &                                           \\ \cline{1-1} \cline{4-4}
\end{tabular}
\end{table}

In this research question, we evaluate whether the centrality-based ranking approach outperforms the statement and branch coverage-based approaches. 
Tables~\ref{tab:stmtcentrality_smtcov_subjects} and Table~\ref{tab:stmtcentrality_branchcov_subjects} show the relative improvement in the percentage of fault detection between the MR sets generated by the Statement Centrality-based prioritization approach and the MR sets generated by a Statement Coverage and Branch-Based Prioritization approaches for BBMap, LingPipe, Linear Regression, and Naive Bayes, respectively. For the BBMap program, Statement Centrality-based prioritization had a higher relative improvement in fault detection effectiveness compared to statement and branch coverage-based prioritization for all the MR set sizes. Similarly, for the Linear Regression program, the Statement Centrality-based approach had higher fault detection effectiveness compared to the Statement and Branch Coverage-based approach for all the MR set sizes. For Naive Bayes, the statement centrality-based approach provided a higher fault detection effectiveness for the MR set size $m=1$ to $m=5$ compared to the statement and branch coverage-based prioritization. Therefore, we reject the null hypothesis $H_{02}$ in general. Overall, our proposed approach provided higher fault detection effectiveness when compared to code coverage-based approach. Based on the result shown in table~\ref{time_method_SUT}, \textbf{our proposed approach shows a reduction in the average time taken to detect a fault ranging from 36\%-93\% relative to the code coverage based approach for the SUT}.\textbf{Answer to RQ2: our proposed statement centrality-based approach provided outperformed the code coverage based approach.}

\subsection{RQ3: Comparison of Statement Centrality-based MR prioritization approach against the Fault-based prioritization approach}


In this research question, we evaluate whether the Statement Centrality-based and Variable-based ranking approaches outperform the Fault-based approaches. 



Table~\ref{tab:stmtcentral_faultbased_subjects} shows the relative improvement in fault detection percentage between the MR sets generated by the Statement Centrality-based prioritization approach and the MR sets generated by Fault-based prioritization approaches for BBMap, LingPipe, Linear Regression, and Naive Bayes respectively. 

For the BBMap program, the Statement Centrality-based method was superior, placing the most effective MR first, leading to better fault detection than the fault-based approach for MR sets sizes 2 to 8. In LingPipe, the fault-based method was less effective, with a 5\% improvement for MR set sizes of 2 to 10 compared to Statement Centrality. For Linear Regression, Fault-based had less than 10\% improvement for MR sizes 1 to 8. Hence, the null hypothesis $H_{03}$ is rejected only for these sizes in Linear Regression. With Naive Bayes, both approaches were equal for MR sizes 4 to 10, not allowing $H_{03}$ rejection. The lack of code coverage diversity in Linear Regression and Naive Bayes led to random MR prioritization. Thus, except for BBMap, Statement Centrality was similar to fault-based. Provided faster fault detection and higher APFD values for BBMap only.\textbf{ Answer to RQ3: Our approach was more effective than the failure-based method only for BBMap.}




\begin{table}[h!]
\centering
\caption{Average relative improvement in fault detection effectiveness using Fault-based approach over Statement Centrality-based prioritization approach}
\label{tab:stmtcentral_faultbased_subjects}
\begin{tabular}{|p{1.2cm}|p{1.2cm} p{1.2cm}|p{1.2cm}|p{1.2cm}}
\hline
\textbf{MR Set-Size} & \multicolumn{1}{c|}{\textbf{BBMap}} & \textbf{LingPipe} & \textbf{Linear Regression} & \multicolumn{1}{c|}{\textbf{Naive Bayes}} \\ \hline
1                    & \multicolumn{1}{c|}{0\%}              & 16.50\%*             & 27.65\%*                      & \multicolumn{1}{c|}{20.48\%*}                \\ \hline
2                    & \multicolumn{1}{c|}{-4.41\%*}          & 3.06\%*              & 3.22\%                       & \multicolumn{1}{c|}{5.26\%*}                 \\ \hline
3                    & \multicolumn{1}{c|}{-4.28\%*}          & 2.81\%*             & 5.97\%*                       & \multicolumn{1}{c|}{1.01\%}                 \\ \hline
4                    & \multicolumn{1}{c|}{-6.94\%*}          & 2.28\%*              & 2.81\%*                       & \multicolumn{1}{c|}{0\%}                    \\ \hline
5                    & \multicolumn{1}{c|}{-4.10\%}          & 1.19\%*              & 12.32\%*                      & \multicolumn{1}{c|}{0\%}                    \\ \hline
6                    & \multicolumn{1}{c|}{-5.26\%*}          & 0.94\%              & 7.89\%*                       & \multicolumn{1}{c|}{0\%}                    \\ \hline
7                    & \multicolumn{1}{c|}{-3.94\%*}          & 0.67\%              & 5.12\%*                       & \multicolumn{1}{c|}{0\%}                    \\ \hline
8                    & \multicolumn{1}{c|}{0\%}              & 0.48\%              & 1.23\%                       & \multicolumn{1}{c|}{0\%}                    \\ \hline
9                    & \multicolumn{1}{c|}{}               & 0.05\%              & 0\%                          & \multicolumn{1}{c|}{0\%}                    \\ \cline{1-1} \cline{3-5} 
10                   & \multicolumn{1}{c|}{}               & 0\%                 & 0\%                          & \multicolumn{1}{c|}{0\%}                    \\ \cline{1-1} \cline{3-5} 
11                   &                                     &                   & 0\%                          &                                           \\ \cline{1-1} \cline{4-4}
\end{tabular}
\end{table}

\section{Related Works}
Authors~\cite{srinivasan2022metamorphic} proposed two MR prioritization approaches: fault-based and coverage-based. These approaches were evaluated on three complex open-source systems and significantly outperformed the current ad-hoc practice in terms of fault detection effectiveness. Sun et al.~\cite{chen2018metamorphic} proposed a path-directed approach to generate and prioritize source test cases for metamorphic testing, demonstrating effectiveness in real-world programs. Mayer et al.~\cite{mayer2006empirical} evaluated factors for selecting good MRs, including fault detection ability, cost, and diversity, finding diversity to be most important for maximizing fault detection. While related work~\cite{chen2018metamorphic}~\cite{mayer2006empirical} focuses on selecting MRs, our work is based on prioritizing MRs.

Chen et al.~\cite{chen2004case} investigated the selection of effective metamorphic relations (MRs) for metamorphic testing (MT). The experiment results suggested that MRs causing greater differences in program executions have a higher failure-detecting capability. The proposed approach focuses on the selection of MRs.

Torres et al.~\cite{duque2023exploring} present MetaTrimmer, a method for selecting and constraining MRs based on test data. The results of the experiment indicate that MetaTrimmer is effective in this task. However, our proposed work focuses on prioritizing MRs. Huang et al.~\cite{huang2022metamorphic} propose a method for selecting and prioritizing MRs based on test adequacy criteria. MRs are sorted using branch coverage Manhattan distance (BCMD) and selected based on their ability to improve test coverage. Experiments show that the proposed method improves defect detection and test effectiveness while using fewer MRs compared to other methods.

Zhou et al.~\cite{zhou2004metamorphic} introduce a method to evaluate the effectiveness of metamorphic relations (MRs) using mutation analysis to prioritize and select MRs. The methodology involves testing ten MR groups and the effectiveness of MRs is evaluated based on whole- and single-mutation scores. The results show that the use of high-priority MRs reduces the number of MRs. The proposed method guides the selection of the MR. Our proposed approach focuses on prioritizing metamorphic relations and does not focus on reducing the number of MRs.

\section{Threat to Validity}
\label{section:ThreattoValidity_method2}

\textbf{External Validity: }

 We evaluated our approach on only four subject programs may not be representative of all software types or application domains. The effectiveness of our approach may vary for applications in other areas. Additionally, while we used standard mutation operators provided by μJava, PIT, and Major, which cover a wide range of code changes, these may not fully represent all possible faults in real-world scenarios.


\textbf{Internal Validity: }

 The effectiveness of our approach may be influenced by the specific set of metamorphic relations used, and different sets of MRs might lead to different results. Additionally, the random baseline and any randomness in our approach (e.g., tie-breaking in prioritization) could introduce variability in results. While we used multiple runs to mitigate this, it remains a potential threat to the internal validity of our conclusions. 

.
\section{Summary}
\label{Section:Conclusion_method2}
In this work, we developed automated approaches for MR prioritization using Statement Centrality information to improve MT efficiency and effectiveness. An empirical study using four complex open-source software systems and machine learning programs showed that our Statement Centrality and Variable-based approach outperforms random and Code Coverage-based approaches, performing closer to Fault-based and Optimal ordering. The proposed approach reduced the fault detection time by 79\% for BBMap and 47\% for LingPipe. 


\bibliographystyle{IEEEtran}
\bibliography{mybib}
\end{document}